\documentclass[12pt, a4paper]{article}

\usepackage{bm}
\usepackage{latexsym}
\usepackage{dcolumn}
\usepackage{amsmath,amsfonts,amssymb}
\usepackage{graphicx,epsfig}
\usepackage{verbatim}
\usepackage{xcolor}
\usepackage[caption=false]{subfig}
\usepackage{slashed}
\usepackage{etex}
\usepackage{float}
\usepackage{mathtools}
\usepackage{mathrsfs}

\usepackage{setspace}
\usepackage[active]{srcltx}
\usepackage{wrapfig}
\usepackage[margin=1in]{geometry}

\def\nn{\nonumber}

\def\l{\left}
\def\r{\right}
\def\DM{\mathrm{d}}
\def\lp{\ell_0}

\def\nn{\nonumber}

\def\l{\left}
\def\r{\right}
\def\DM{\mathrm{d}}
\def\lp{\ell_0}

\def \lp {L_0}

\def \lp {\ell_0}

\def \Rsq {\widetilde {\rm \bf Ric}(p;p_0)}
\def \Rsqn {\widetilde {\rm \bf Ric}}
\def \Rsg {{\rm \bf Ric}(p_0)}
\def \Rsgp {{\rm \bf Ric}(p)}

\usepackage{empheq}

\begin{document}

\begin{center} {\Large \bf
Limits of a non-local quantum spacetime
}
\end{center}

\bigskip

\centerline{Dawood Kothawala}

\vspace{0.25cm}

\medskip
\centerline{Department of Physics}
\centerline{Indian Institute of Technology Madras} 
\centerline{Chennai 600 036, India}

\vspace{0.25cm}

\centerline{email: dawood@iitm.ac.in}

\vspace{0.05in}

\centerline{(submitted on 20 March, 2023)}

\bigskip

\begin{abstract}
\noindent 
A generic implication of incorporating gravitational effects in the analysis of quantum measurements is the existence of a zero-point length of spacetime. This requires an inherently non-local description of spacetime, beyond the usual one based on metric $g_{ab}(x)$ etc. The quantum spacetime should instead be reconstructed from non-local bi-tensors of the form $\mathscr{G}_{ab \ldots i'j' \ldots}(x,x')$. A deeper look then reveals a subtle interplay interplay between non-locality and the limit $G\hbar/c^3 \to 0$. In particular, the so called emergent gravity paradigm --  in which gravitational dynamics/action/spacetime are emergent and characterised by an {\it entropy functional} -- arises as the \textit{Cheshire} grin of a fundamentally non-local quantum spacetime. This essay describes the flow of metric with respect to Planck length, and proposes a novel action for the same.
\\
\begin{center}
\textit{Dedicated to T. Padmanabhan} -- ``{\it Paddy}".
\end{center}
\end{abstract}


\begin{center}
\line(1,0){300}
\end{center}

\begin{center}
\bf Essay written for the Gravity Research Foundation 2023 Awards for Essays on Gravitation.
\end{center}

\pagestyle{empty} 

\pagebreak

\pagestyle{plain}
\setcounter{page}{1}

\begin{flushright}
{\footnotesize{
\it The question of the validity of the hypotheses of geometry 

in the infinitely small is bound up with the question of the

ground of the metric relations of space \ldots
}

- Riemann (1854)
}
\end{flushright}

\noindent \textbf{\textit{Limits of spacetime}}
\\
\hspace{1cm} \textit{\ldots Planck length and non-locality}

\vspace{.25cm}

Suppose a civilisation (ours or otherwise) has eventually discovered the complete framework of quantum gravity, and is looking to extract the good old laws of general relativity given by Einstein as a limit of their framework. Are they assured to recover, say, the conventional action of gravity -- the Einstein-Hilbert action? If not, what \textit{will} the limit give? These questions might not matter to the civilisation that has the complete laws of quantum gravity in hand, but they can serve as important guideposts to the one that is nowhere close to it. Singular limits abound in nature, and if the limit mentioned above is singular, our fundamental description of classical gravity and spacetime itself might need a revision. Consider the following story involving (yet again!) the apple, by Michael Berry \cite{berry}: ``\textit{Biting into an apple and finding a maggot is unpleasant enough, but finding half a maggot is worse. Discovering one-third of a maggot would be more distressing still: The less you find, the more you might have eaten. Extrapolating to the limit, an encounter with no maggot at all should be the ultimate bad-apple experience. This remorseless logic fails, however, because the limit is singular: A very small maggot fraction $(f \ll 1)$ is qualitatively different from no maggot $(f = 0)$.}" 
In this essay, I will argue that something similar happens for gravity and spacetime, except in this case, the (non-)existence of the worm leaves its imprint through certain well established clues that have already been given to us! 
The issue of taking limits of a spacetime with respect to constant parameters that characterize it (eg. total mass $M$) must be handled with care, as Geroch observed long back in his classic paper, Ref. \cite{geroch}. The issue acquires fundamental importance if spacetime is inherently non-local with a fundamental length scale, say $\lp$. What, then, would be its limit when $\lp \to 0$? In this essay, I will first argue that spacetime, even classically, must be described in terms of non-local observables, and then incorporate $\lp$ in such a description to address the above question.

\vspace{.75cm}

\noindent \textbf{\textit{Reconstructing spacetime from observational tools}}
\\
\hspace{1cm} \textit{\ldots replacing metric $g_{ab}(x)$ with bi-tensors $\mathscr{G}(x,y)$}

\vspace{.25cm}

\begin{figure}[H]
	\begin{center}
	\scalebox{0.4}{\includegraphics{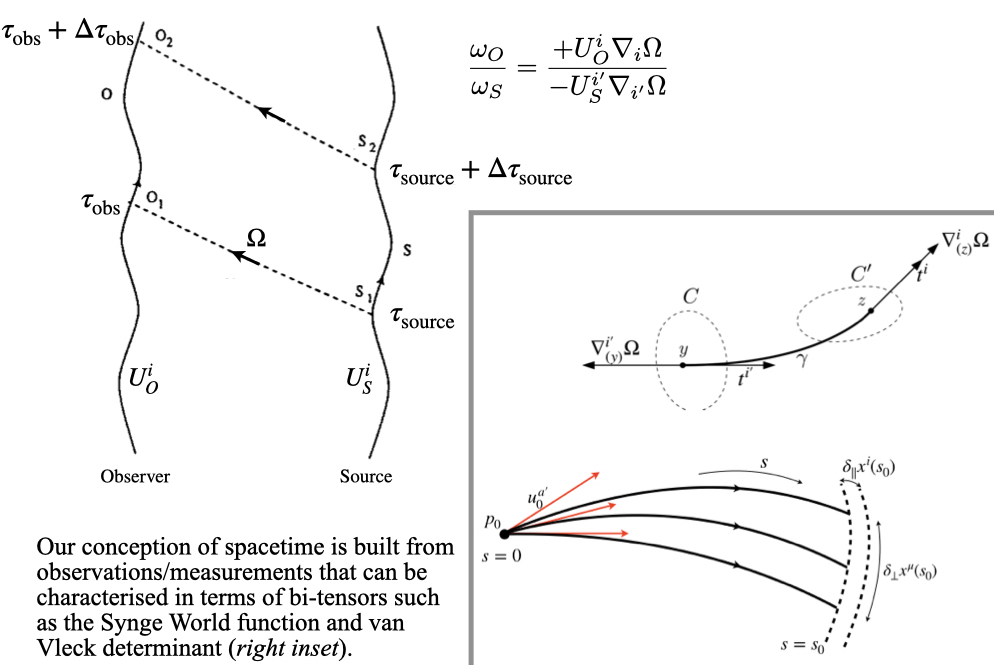}}
	\end{center}
	    \caption{Reconstructing spacetime from measurements.}%
	    \label{fig:synge-wfn-measurements}%
	\end{figure}

Our conception of space and time is built through measurements and observations of physical phenomenon, characterised and catalogued in terms of extended solid objects and light rays -- ``{\it rods and clocks}" -- which we employ as probes; see \textbf{Fig \ref{fig:synge-wfn-measurements}}.

\noindent It is but natural to then expect that the quantum behaviour of these probes should play an important role in characterising the structure of spacetime. In brief:

\vspace{.25cm}
\begin{flushleft}
\textcolor{black}{$\spadesuit$} \hspace{.15cm} \textsf{Spacetime should be re-constructed purely from quantities characterising measurements.}
\end{flushleft}
\vspace{.25cm}

\noindent While this basic, fundamental fact was emphasized by Einstein on the very first page of his foundational paper on {\it Special Relativity}, it has been forgotten through the long history of gravitational physics. As a result, Einstein's great insight to describe gravity in terms of geometry of space and time got firmly rooted into a description of gravitational dynamics using local, tensorial quantities such as $g_{ab}(x), R_{abcd}(x), etc.$. While these quantities can, and do, characterise motion and measurements, they miss the essence of the fact highlighted above: spacetime must be re-constructed from observables tied directly to measurements. This shift of viewpoint, which might seem to be a matter of taste at the classical level, becomes absolutely essential at the quantum level where local, tensorial objects might not make much sense. In fact, if the quantum effects are non-analytic and non-local, then the limit in which one recovers locality might not commute with the limit $\hbar \to 0$, and hence: 
\vspace{0.25cm}
\begin{flushleft}
\textcolor{black}{$\spadesuit$} \hspace{.15cm} \textsf{The classical limit might carry a vestige of an inherently quantum spacetime.}
\\
\end{flushleft}
\vspace{0.25cm}
Nowhere is this statement more beautifully, and more succinctly, captured than in the prescient remarks of Riemann (quoted above) from his classic lecture on the foundations of geometry. As Riemann prophesizes, our usual conceptions of space might not hold at very small scales. And so it goes for gravity. 

\vspace{.25cm}

\noindent In this essay, I will describe how one can, and should, re-construct spacetime in terms of non-local bi-tensors, $\mathscr{G}_{ab \ldots i'j' \ldots}(x,x')$ that depend on two points. Specifically, I will first describe how classical metric can be described in terms of such quantities. When there exists a fundamental length scale $\lp$ in the theory, $\mathscr{G}$ can depend on $\lp$, i.e. $\mathscr{G} \equiv \mathscr{G}_{ab \ldots i'j' \ldots}(x,x'; \lp)$, through the dimensionless ratio of geodesic interval between $x$ and $x'$ and $\lp$, with a non-trivial $\lp \to 0$ limit. I demonstrate this by obtaining an ``effective metric" for the quantum spacetime by imposing the condition that there exists a lower bound on spacetime intervals. As a bonus, we will then look closely at the structure of the gravitational action built from this effective metric, and show that, even classically, the local limit of this action is not given by the Einstein-Hilbert lagrangian. Instead, it is given by the so called {\it entropy functional} of the emergent gravity paradigm, thereby leading us to the surprising conclusion that forms the title of this essay:
\vspace{0.25cm}
\begin{flushleft}
\textcolor{black}{$\spadesuit$} \hspace{.15cm} \textsf{Emergent gravity is a relic of a quantum spacetime.}
\\
\end{flushleft}
%

\vspace{.75cm}

\noindent \textbf{\textit{Echoes from the mesoscopic spacetime}} 
\\
\hspace{1cm} \textit{\ldots going down the rabbit hole}

\vspace{.25cm}
 
The conclusion above would connect two of the most robust, and powerful, clues that have been obtained when trying to combine the basic principles of quantum mechanics and general relativity, and are expected to be survive any eventual overhauls which a complete theory of quantum gravity might bring in. These are: 

\indent \;\; \textit{Clue 1}: Gravitational effects in measurements by quantum probes, and \\
\indent \;\; \textit{Clue 2}: Thermal aspects associated with Rindler (acceleration) horizons 

\noindent Clue 1 has been derived in a large number of ways, starting right from a variant of Heisenberg's microscope extended to incorporate gravitational effects, to the analysis of operational probes such as clocks and identify their limitations in measuring time accurately. The literature on this topic is vast, but the essence of all the results can be extracted into a quantitative statement: If Lorentz invariance is kept intact, then effects of a quantum spacetime with a zero point length can be captured by demanding that the distance function satisfies
$$
d(x, x) \neq 0
$$
which violates the \textit{identity of the indiscernibles}. An event in spacetime can not be truly discernible due to quantum fluctuations which prohibit localisation with a precision better than $\lp$. Of course, a distance function that captures this fact must be thought of as an effective quantum object, and will not conform to the standard axioms of metric spaces. Even the equality part of triangle inequality will not hold, since an event can not be localised with a precision better than $\lp$. As for Clue 2, perhaps it's most significant impact has been the development of the so called \textit{Emergent gravity} paradigm, set along the line of arguments that can succinctly be described as: 
	\begin{figure}[H]
	\begin{center}
	\scalebox{0.45}{\includegraphics{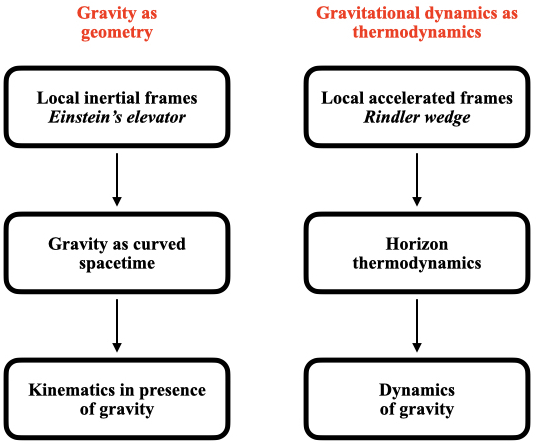}}
	\end{center}
	    \caption{Spacetime thermodynamics and Emergent gravity.}%
	    \label{fig:rindler-emergent}%
	\end{figure}
\noindent \textbf{Fig \ref{fig:rindler-emergent}} covers the spirit of all the emergent gravity paradigms. As we go deeper down the rabbit hole, these paradigms sort themselves into the \textit{taxonomy of emergence}, depending on what emerges: gravitational dynamics, gravitational action, or spacetime itself. In what follows, we will follow the rabbit hole back up, starting from the the small scale structure of spacetime, and employing the right set of tools to describe it the emergence of a mesoscopic description of spacetime, gravitational action, and gravitational dynamics. 

\vspace{0.25cm}

\noindent Coming out from the rabbit hole, we now stare back at what we recognise as a surprising relic of the small scale structure of spacetime that has survived the journey back to $\hbar=0$.

\vspace{.75cm}

\noindent \textbf{\textit{Synge's World function and the architecture of spacetime}} 
\\
\hspace{1cm} \textit{\ldots locality without locality}

\vspace{.25cm}

Differential geometry already provides us with objects that can be used to characterise spacetime in the light of the discussion above. And the most important such object - the one that will pervade all throughout this essay - is Synge's World function $\Omega(x, y)$ and the van Vleck determinant $\Delta(x, y)$. The World function is nothing but one half the (signed) square of geodesic interval between pairs of spacetime events, while $\Delta(x, y)$ measures spread of geodesics emanating from the point $x$ and reaching points $y$ at fixed geodesic distance from $x$ (see inset in \textbf{Fig \ref{fig:synge-wfn-measurements}}). Usually derived from the metric, our final aim will be to introduce it as more fundamental than the metric, and incorporate features of the quantum spacetime through $\Omega(x, y)$. Metric tensor thus becomes a derived object, and will in particular not have the mathematical features expected of it when quantum corrections are introduced. The curvature, action, etc built from such an effective metric will be our main focus.

\noindent Let me first describe how, even in \textit{classical} geometry, one can trade-off metric, curvature etc for a more economical description in terms of $\Omega(x, y)$. The possibility of doing so is based on the existence of mathematical identities relating the local tensorial quantities (metric curvature, etc.) that characterise a spacetime to the coincidence limit (denoted by ``$[\ldots]$") of derivatives of $\Omega(x,y)$. For example, we have, in the limit $x \to x'$
	\begin{eqnarray}
	g_{a'b'} = g_{ab} &=& \l[{\nabla}_a {\nabla}_b \Omega(x,x') \r] = \l[{\nabla}_{a'} {\nabla}_{b'} \Omega(x,x') \r] 
	\nn \\
	R_{a' (c' d') b'} &=& \l(3/2\r) \l[{\nabla}_a {\nabla}_b \nabla_{c} \nabla_{d} \Omega(x,x') \r]
	\end{eqnarray}

\noindent One can now use Clue 1 concerning zero-point length to demand the following of the \textit{quantum} spacetime: (1) Geodesic distances, and hence $\Omega(x,y)$, have a Lorentz invariant lower bound, say $\lp^2$; we quantify this by modifying distances $\sigma^2 \rightarrow {\sf S}_{\lp} \l[ \sigma^2 \r]$ such that ${\sf S}_{\lp} \l[ 0 \r] = \lp^2$.
%
where $\sigma^2=2 \Omega$ is the squared geodesic interval. Precise details of the function ${\sf S}_{\lp} \l[ \sigma^2 \r]$ must come from a complete framework of quantum gravity, and we will not make any assumptions about it here. The second input is of a somewhat technical nature, but can be simply motivated by the demand that the zero point length also acts as a UV regulator of two point functions $G(x, y)$ (see, for instance, \cite{dewitt}). Since typically the two point functions have the singular structure given by the Hadamard form  	
	\begin{eqnarray}
	G(x,y) := \frac{\Delta^{1/2}}{\sigma^{D-2}} \times \l( 1 + \mathrm{subdominant~terms} \r)
	\end{eqnarray}
it is then evident that the reconstructed quantum metric will depend on the bi-tensors $\Delta(x,y)$ and $\sigma(x,y)^2$. Using the mathematical identities associated with these bi-tensors, the qmetric can now be obtained as \cite{qmetric, pesci-grg-review}.
	\begin{eqnarray}
	{\textsf q}_{ab}(x;y) &=& {\sf A} \; g_{ab}(x) - \epsilon\l( {\sf A} - \frac{\sigma^2 {\sf S}'^2}{\sf S} \r) \; t_a(x;y) \; t_b(x;y)
	\nn \\
	{\sf A} &=& \frac{{\sf S}_{\lp}}{\sigma^2} \l(\frac{\Delta_{\phantom{\mathcal S}}}{\Delta_{\sf S}}\r)^{+\frac{2}{D_1}}
	 \label{eq:qmfinal}
	\end{eqnarray}
where $t_a$ is the tangent vector to the geodesic connecting $y$ with $x$. This is the final form of the effective quantum metric - the \textit{qmetric} - which will be the key object that characterises the quantum spacetime. It has the following key mathematical properties:

\noindent	$\star$ ${\textsf q}_{ab}(x;y)$ is a \textit{non-local bi-tensor}, determined by the geodesic structure of the spacetime.

\noindent	$\star$ $\lim \limits_{\lp \to 0} {\textsf q}_{ab}(x;y) = g_{ab}(x)$, while ${\textsf q}_{ab}(x;y)$ is \textit{singular} in the limit $\sigma^2 \to 0$.
\\
\noindent These features lead to several interesting and subtle insights, largely related to the nature of the interplay of the limits $\sigma^2 \to 0$ and $\lp \to 0$; in particular, to recover the classical results, we have the following two candidate limits
\begin{eqnarray}
\lim \limits_{\lp \to 0} \lim \limits_{y \to x} 
\;\;\; {\rm and} \;\;\; 
\lim \limits_{y \to x} \lim \limits_{\lp \to 0}
\label{eq:lims}
\end{eqnarray}
\noindent -- a situation not encountered in the local description in terms of $g_{ab}(x)$ etc. If one fixes the event $y$, ${\textsf q}_{ab}(x;y)$ becomes a standard second rank tensor at $x$, coupled disformally to $g_{ab}(x)$. One may then do standard differential geometry using this metric, and in particular probe the $\lp \to 0$ limit of ``{local scalars}" constructed out of $q_{ab}$. 

\vspace{.75cm}

\noindent \textbf{\textit{Emergent gravity emerges as a relic of the quantum spacetime}}
\\
\hspace{1cm} \textit{\ldots back up the rabbit hole, the ``zero-point length" leaves it's Cheshire grin!}

\vspace{.25cm}

\noindent The simplest object to focus on is the Einstein-Hilbert lagrangian, given simply by the Ricci ``scalar", and ask:
	\begin{eqnarray}
	  \underset{p \rightarrow p_0}{\lim} \Rsqn (p;p_0) \overset{?}{=} \Rsg + {\rm terms~of~order~} \lp
	\nn
	\end{eqnarray}
If the answer to this is yes, it would imply that local limit of the Ricci (bi-)scalar of the quantum spacetime has an expansion in $\lp$ whose leading term is the Ricci scalar of $g_{ab}(x)$. If the limits highlighted in Eq. (\ref{eq:lims}) were the same, this would indeed be the case. \textit{This, however, turns out not to be the case!} In fact, not only do we have $${\lim \limits_{\lp \to 0} \lim \limits_{\sigma^2 \to 0}} \Rsq \neq \Rsg$$ but what we do get for these limits turns out to have important implications for the emergent gravity paradigm (\cite{EG}). An exact (though painfully lengthy) computation instead yields
\begin{eqnarray}
	 \underset{p \rightarrow p_0}{\lim} \Rsq
	= \underbrace{\alpha \l[R_{ab} t^a t^b \r]_{p_0}}_{O(1) \; \rm term} 
\; + \;
{\lp^2} \;
\underbrace{
\l[~\rm curvature~squared~terms~ \r]_{p_0}
}_{O(\lp^{\;2}) \; \rm term}
\end{eqnarray}
where $\alpha$ is a numerical constant (and depends on spacetime dimensions). 
Thus, although $q_{ab} \to g_{ab}$ when $\lp=0$, curvature scalars generically do not! 

\begin{figure}[H]
	\scalebox{0.5}{\includegraphics{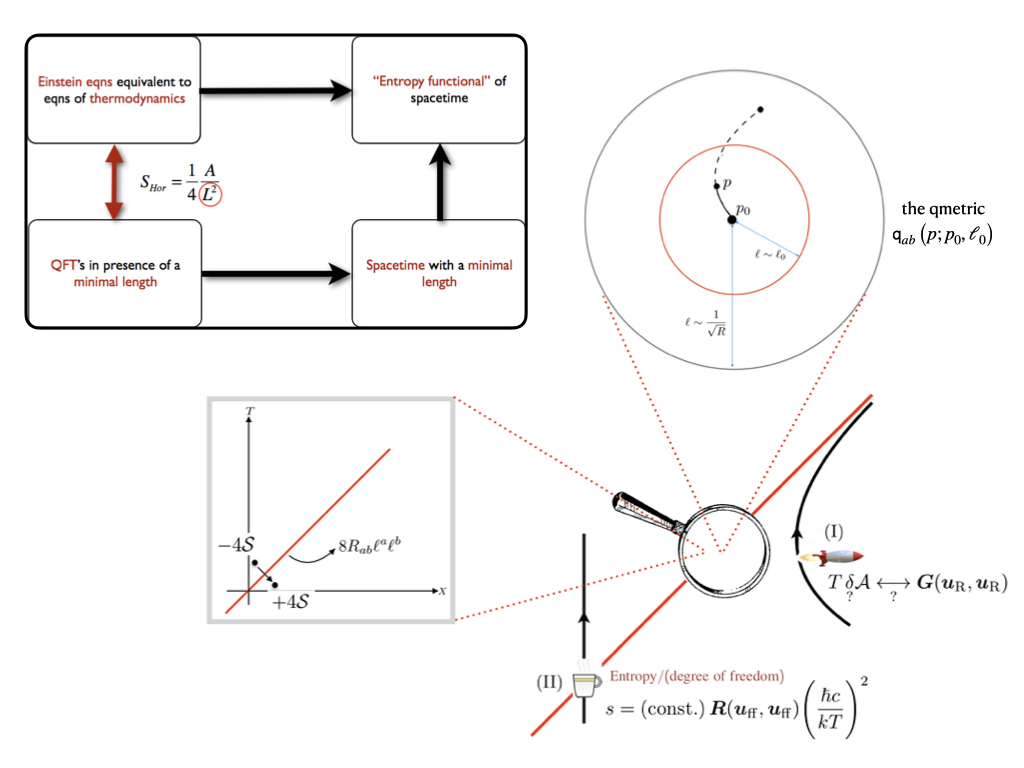}}
	   \caption{Small scale structure of spacetime, thermodynamics, and emergent gravity.}%
	   \label{fig:ml-eg}%
	\end{figure}

\noindent The cognoscenti will immediate recognise the connection with emergent gravity. The leading term above is precisely the entropy density $R_{ab} q^a q^b$ that determines the entropy functional of the emergent gravity paradigm. From there, one proceeds to vary the vectors $q^a$ and obtain Einstein equations as a constraint. This is an extremely non-trivial and important result, since we have not used any aspect of horizon thermodynamics that usually is the starting point of all ideas pertaining to emergent/entropic gravity. Why then do we get such a result? I would like to suggest that this is closely related to the following fact: introducing a non-trivial quantum mechanically induced non-locality helps us go beyond the classical thermodynamics of horizons, and introduces some features associated with the statistical mechanics of more fundamental degrees of freedom. In the mesoscopic domain, these degrees of freedom are effectively represented by $\Omega(x, y)$ and its concomitants. Mathematically, two key ideas in emergent gravity -- {\it local Rindler frames} as probes of spacetime curvature (due to Jacobson), and a variational principle based on {\it entropy functional} (due to Padmanabhan et. al.) -- find a unified and purely geometric description in our framework in terms of equi-geodesic {\it surfaces} $\Omega(x, {\rm fixed}\; y)=$ const (which replace the Rindler {\it trajectories}) and the $\lp=0$ term of the coincidence limit of Ricci bi-scalar of the qmetric, which happens to have the same form as the entropy functional. The upper left block in \textbf{Fig \ref{fig:ml-eg}} summarizes the reconnections described above. 
\noindent The key theme of this essay also addresses one of the most frequently asked questions about the emergent gravity paradigm: \textit{Why choose it over the conventional general relativity based on the Einstein-Hilbert lagrangian}? Most of our conventional intuition about quantum gravity is built on the idea that classical gravity can be obtained as a ``Taylor series expansion" in $\lp^2$ starting from quantum gravity. This, in turn, assumes that all quantum gravitational effects are analytic in $\lp^2$ and will lead to some sensible classical limits when
$\lp \to 0$ limit is taken. Some thought shows that this is a highly questionable assumption and we should take seriously the possibility that quantum gravity
could have features which are non-analytic in $\lp$. In that case, the process of taking limits might involve manipulating singular quantities leading to unexpected (but interesting) results. Our results imply that this is indeed what happens in the case of gravity. Similar issues were also emphasized by Brown in \cite{brown}. 

\vspace{.75cm}

\noindent \textbf{\textit{Epilogue}}
\\
\hspace{1cm} \textit{a non-local action for gravity \ldots and the roads ahead}

\vspace{.25cm}

Although we have so far dealt with limits, either in $\lp$ or $\sigma^2$, the qmetric does yield a closed form expression for the Ricci bi-scalar $\Rsq$, and we must ask what a more fundamental action built from this object might look like, without resorting to any expansion in any of the parameters. Such an action must be evidently non-local, and one can only speculate as to it's structure. I propose the following construction: Consider any event $p$ at which you construct the qmetric anchored on an event $p_0$ (thus it is a scalar at $p_0$). Compute the integral of 
$\Rsq$ over all $p_0 \in I^-(p)$ - the causal past of $p$ - with measure $\DM \mu = \DM v(p_0)/v^-(p)$, where $\DM v(p_0)$ is a local volume measure, and $v^-(p)$ is the volume of $I^-(p)$ with respect to this measure, and ensures the normalization $\int_{p_0 \in I^-(p)} \DM \mu(p,p_0) = 1$. Integrate the final result over all $p$. Mathematically,
\begin{eqnarray}
{\rm Action} = \int \limits_{{\rm all} ~ p} \DM v(p) \int \limits_{p_0 \in I^-(p)} \Rsq \; \DM \mu(p,p_0)
\end{eqnarray}
Note that, when $\lp=0$, 
$\Rsq = \Rsgp$, and we recover the standard Einstein-Hilbert action: $\int_{{\rm all} ~ p} \DM v(p) \Rsgp$. On the other hand, keeping $\lp$ non-zero, the contribution from events $p_0$ ``close" to $p$ would give some weighted integral of the entropy functional. At intermediate scales, the above action would generically have a complicated structure. More so when one proceeds to vary it. Given that the background metric $\bm g$ appears only as a convenient auxiliary field, while the bi-scalars introduced so far play the more fundamental role, it is these objects which one should vary to extremize the action. Although a complete understanding of such a variational principle, and even its mathematically rigorous implementation, would require much greater effort. It should nevertheless be amenable to a combination of analytic and numerical methods. 

\noindent I expect the ideas and results described in this essay -- summarised in {\bf Fig \ref{fig:ml-eg}} -- will contribute to a better understanding of, and a clear distinction between, the essential and non-essential features of any candidate theory of quantum gravity, features 
ensure that certain robust results of the semiclassical theory are not only recovered, but find a better explanation. Emergent gravity is but one such feature; the qmetric says much more about several other features of the mesoscopic spacetime. For instance, geometrical structures described by the qmetric imply an effective dimensional reduction at small scales \cite{sc-tp-dk}. Some preliminary results indicate that the spacetime singularities will also be regularised \cite{sc-dk-pesci, dk-singularities}. Moreover, the causal structure associated with the qmetric implies ``silence" at small scales \cite{dk-singularities}.

\noindent A deeper study of these features, treading clear of fashion and trends, might imply a significant change in perspective at the classical level itself. This requires a judicious choice of tools to start with, and explore their implications, which was the main focus of this essay. We hope that future work along these lines will reveal more important results that can serve as guideposts enroute to a complete theory of quantum gravity.

\pagebreak


\end{document}